\documentclass[aps,prl,floatfix,twocolumn,showpacs,10pt,longbibliography]{revtex4-1}
\usepackage{amsmath}
 \usepackage{relsize}
\usepackage{url}
\usepackage{graphicx}
\usepackage{dcolumn}
\usepackage{bm}
\usepackage{amssymb}
\usepackage{rotating}
\usepackage[abs]{overpic}
\usepackage{xcolor}
\usepackage{tabularx}
\usepackage{floatrow}
\usepackage{subfig} 
\usepackage[justification=RaggedRight]{caption}
\usepackage{hyperref}
\hypersetup{colorlinks = true, citebordercolor={black}, linkcolor={black}, citecolor={black}, urlcolor={black}}
\begin{document}



\title{Potential Coexistence of Exciton and Fermion Pair Condensations}

\author{LeeAnn M. Sager, Shiva Safaei, and David A. Mazziotti}

\email{damazz@uchicago.edu}

\affiliation{Department of Chemistry and The James Franck Institute, The University of Chicago, Chicago, IL 60637}%

\date{Submitted July 13, 2019\textcolor{black}{; Revised January 22, 2020}}





\begin{abstract}
Extensive theoretical and experimental investigation has been conducted on fermion pair condensation and exciton
condensation as distinct classes of Bose-Einstein-like condensation.  In this work, the existence of a fermion-exciton
condensate---a single quantum state in which character of both fermion pair and exciton condensates coexist---is
established computationally in the low-particle-number ($N$) limit \textcolor{black}{and theoretically in the
large-$N$ thermodynamic limit.}  The trade-off between the fermion pair and excitonic character of the fermion-exciton
condensate is shown to be elliptic in nature.  The possibility that the properties of fermion-exciton condensates
could be a hybrid of the properties of fermion pair condensates and exciton condensates is discussed, and future
experimental and computational exploration of this new class of condensate\textcolor{black}{, which may
potentially be realizable in a bilayer of superconductors,}  is anticipated.
\end{abstract}

\maketitle

\textit{Introduction:} Ample experimental and theoretical investigation has centered around \color{black} the condensation of fermion pairs [\onlinecite{BCS1957, Blatt_SC,Anderson_2013, Drozdov_250}] and excitons [\onlinecite{KSE2004,TSH2004,Fil_Shevchenko_Rev,Shiva,Kogar2017,LWT2017,varsano_2017}]. \color{black}  Fermion pair condensates---the most familiar of which include the class of Bardeen-Cooper-Schrieffer (BCS) superconductors [\onlinecite{BCS1957}]---occur when \color{black}particle-particle \color{black} pairs condense into a single quantum state to create a superfluid.  For condensates of Cooper (electron) pairs, the superfluidity of the electrons cause the material through which they flow to be both a perfect conductor and a perfect diamagnet [\onlinecite{Anderson_2013}].  Similarly, exciton condensates involve the condensation of particle-hole pairs (excitons) into a single quantum state to create a superfluid \color{black} associated with the nondissipative transfer of energy [\onlinecite{Fil_Shevchenko_Rev}, \onlinecite{keldysh_2017}]. \color{black} Exciton condensates have been experimentally observed in optical traps with polaritons [\onlinecite{KRK2006,BKY2014,DWS2002,BHS2007}] and the electronic double layers of semiconductors [\onlinecite{SEP2000,KSE2002,KSE2004,TSH2004,NFE2012}] and graphene [\onlinecite{LWT2017}, \onlinecite{Lee2016}-\onlinecite{Li2016}].

In order to combine the frictionless transfer of electrons of fermion pair
condensates and the frictionless transfer of energy of exciton pair condensates, it may be beneficial to explore a
system composed of both fermion pair and excitonic condensations.
Both condensates are known to exist in systems designed to use exciton condensates to mediate the creation of Cooper
pairs at higher
temperatures [\onlinecite{LTSK2012}, \onlinecite{SCKP2018}].
However, this coexistence
of fermion pair and excitonic condensation occurs in two adjacent systems that interact with one another (such as a
superconducting ring deposited around a semiconductor microcavity [\onlinecite{SCKP2018}]) instead
of existing in a joint fermion-exciton condensate state.
As such, the properties of each condensate exist separately from one another instead of
creating a system with the combined properties of both.  \textcolor{black}{As fermion pair condensation and
exciton condensation are known to exist in superconductors and bilayer systems, respectively, a more
promising prospective avenue for obtaining this coexistence may be in bilayer systems constructed from
superconducting layers}.

\begin{figure}[tbh!]
  \includegraphics[trim=15 0 0 0,clip,width=8.5cm,angle=0]{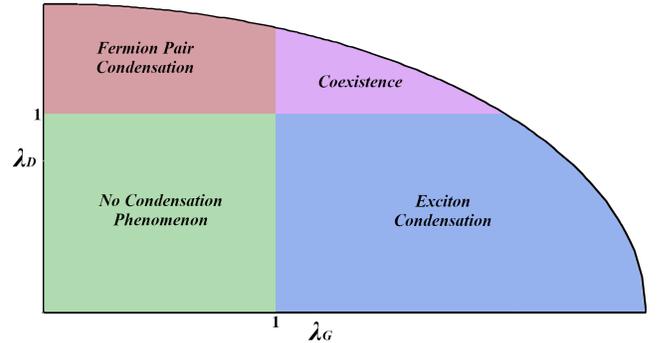}
  \caption{\color{black} A figure demonstrating the elliptic trade-off between the signatures of fermion pair condensation, $\lambda_D$, and exciton condensation, $\lambda_G$, is shown. \color{black}}
  \label{fig:Tradeoff}
\end{figure}

In this study, we address the possible coexistence of a fermion-exciton condensate that contains both fermion pair condensation
and exciton condensation in a single quantum state (see Fig. \ref{fig:Tradeoff}).
To this end, we use the theoretical signatures of fermion pair condensation---independently discovered by Yang [\onlinecite{Y1962}]
and Sasaki [\onlinecite{S1965}]---and exciton condensation---derived by Garrod and Rosina [\onlinecite{Shiva}, \onlinecite{GR1969}]---to
explore the fermion pair and exciton character of few-particle systems.  \textcolor{black}{Furthermore, we prove that a large class of
fermion-exciton condensate wavefunctions  can be constructed by entangling any fermion-pair condensate wavefunction with any
exciton-condensate wavefunction, from which we establish the dual condensate's existence in the large-$N$ thermodynamic limit.}

\textit{Theory:} Bosons are able to condense into
a single, lowest-energy orbital.  A signature of this so-called Bose-Einstein condensation is a large eigenvalue of
the one-boson reduced density matrix [\onlinecite{Penrose_BEC}]  \color{black} with elements \color{black} given by
\color{black}
\begin{equation}
^{1} D^i_j = \langle \Psi |{ \hat b}^{\dagger}_i {\hat b}_j | \Psi \rangle
\label{eq:D1}
\end{equation}
\color{black}
where $|\Psi\rangle$ is defined to be \color{black} a \color{black} $N$-boson wavefunction, each number represents both spatial and spin components of
the boson, \color{black} $i, j$ correspond to one-boson orbitals in a finite basis set with rank $r$, \color{black} and \color{black}${\hat b}^{\dagger}$ and ${\hat b}$ \color{black} are bosonic creation and annihilation operators respectively.

Fermions, however, must obey the Pauli exclusion principle.  Hence, the occupation of a given
spin orbital must be either zero or one, and the one-fermion reduced \color{black} density \color{black} matrix (1-RDM) must have eigenvalues bounded
above by one.  For fermion condensation to occur, pairs of fermions (creating a bosonic state) must condense
into a single two-electron function [\onlinecite{Y1962,C1963,S1965,RM2015}]\color{black}; this two-electron analogue to the one-electron orbital is referred to as a geminal [\onlinecite{Srev_1999, shull_1959}].\color{black}  The signature of fermion condensation is hence related to the \color{black} two-fermion reduced density matrix (2-RDM) with elements \color{black} given by
\color{black}
\begin{equation}
^{2} D_{k,l}^{i,j} = \langle \Psi | {\hat a}^{\dagger}_i {\hat a}^{\dagger}_j {\hat a}_k {\hat a}_l  | \Psi \rangle
\label{eq:D2}
\end{equation}
\color{black}
where $|\Psi\rangle$ is \color{black} a \color{black} $N$-fermion wavefunction, each number represents both spatial and spin components of the fermion, \color{black} $i, j, k, l$ correspond to one-fermion orbitals in a finite basis set with rank $r$, \color{black}  and
\color{black} ${\hat a}^{\dagger}$ and ${\hat a}$ \color{black} are fermionic creation and annihilation operators respectively. In fact, Yang [\onlinecite{Y1962}]
and Sasaki [\onlinecite{S1965}] have independently demonstrated that a large eigenvalue of the 2-RDM (above the bound of one from the
Pauli exclusion principle) is a signature of fermion pair condensation.  Additionally, Sasaki [\onlinecite{S1965}]
has proven that the eigenvalues of the 2-RDM are bounded by $N$ for systems of $2N$ or $2N+1$ fermions in the limit of
strong correlation.

Analogous to fermion-fermion condensation into a single particle-particle function, exciton condensation is the
condensation of particle-hole pairs (excitons) into a single particle-hole function.  By comparison,
one may hence expect a signature of
exciton condensation to be a large eigenvalue in the particle-hole RDM [\onlinecite{Shiva}, \onlinecite{GR1969}, \onlinecite{Kohn1970}] \color{black} with elements \color{black} given by
\color{black}
\begin{equation}
^{2} G_{k,l}^{i,j} =  \langle \Psi | {\hat a}^{\dagger}_i {\hat a}_j {\hat a}^{\dagger}_l{\hat a}_k  | \Psi \rangle
\label{G2}
\end{equation}
\color{black}
similar to the large eigenvalue of the fermionic 2-RDM for fermion pair condensation.  However, there are two large eigenvalues
of the particle-hole RDM
with one corresponding to a ground-state-to-ground-state projection, not excitonic condensation [\onlinecite{GR1969}].  To eliminate
the extraneous large eigenvalue, we construct a modified particle-hole matrix with the ground-state resolution removed
\color{black} with elements
\begin{equation}
^{2} {\tilde G}_{k,l}^{i,j} =^2G_{k,l}^{i,j} -{^{1}D_{k}^{i}} {^{1} D_{l}^{j}}
\label{G2tilde}
\end{equation}
\color{black}
which we denote as the $^2\tilde{G}$ matrix.  While in the noninteracting limit, the eigenvalues of the $^2\tilde{G}$
matrix are zero or one, Garrod and Rosina have shown that the eigenvalue of the
$^2\tilde{G}$ matrix is bounded by $\frac{N}{2}$ for an $N$-electron system in the limit of strong correlation [\onlinecite{GR1969}].  This bound also
describes the maximum number of excitons in a condensate.

\textit{Results:} \textbf{Unconstrained Calculations---}In order to determine whether there is possible coexistence of exciton character and fermion pair character,
\color{black}  general $N$-fermion wavefunctions in $r=2N$ orbitals were constructed, and the coefficients of the wavefunctions
 were
then optimized with respect to the signatures of fermion pair condensation ($\lambda_D$) and exciton condensation ($\lambda_G$).
Specifically, \color{black}
multiobjective optimization was performed \color{black} on $|\Psi\rangle$ \color{black} with respect to a variable $\lambda_{DG}$, which depends on the largest eigenvalues of the $^2\tilde{G}$ matrix ($\lambda_G$) and the $^2D$ matrix ($\lambda_D$) according to
\begin{equation}
\lambda_{DG}=w_G(\lambda_G-\lambda^o_G)^2-(1-w_G)\lambda_D
\label{funcDG}
\end{equation}
where $w_G$ describes the weight given to the optimization of the largest eigenvalue of $^2\tilde{G}$ to the specified eigenvalue provided ($\lambda^o_G$) and where $\lambda_D$ is left unconstrained.  This optimization was conducted through use of a sequential quadratic programming (SQP) algorithm [\onlinecite{maple_2019}, \onlinecite{Gill_SQP}] with gradients computed by second-order centered finite differences [\onlinecite{Fornberg_CFD}].

\begin{figure}[tbh!]
  \centering
  \sidesubfloat[N=3]{\label{fig:N3}\includegraphics[scale=.32,angle=-90]{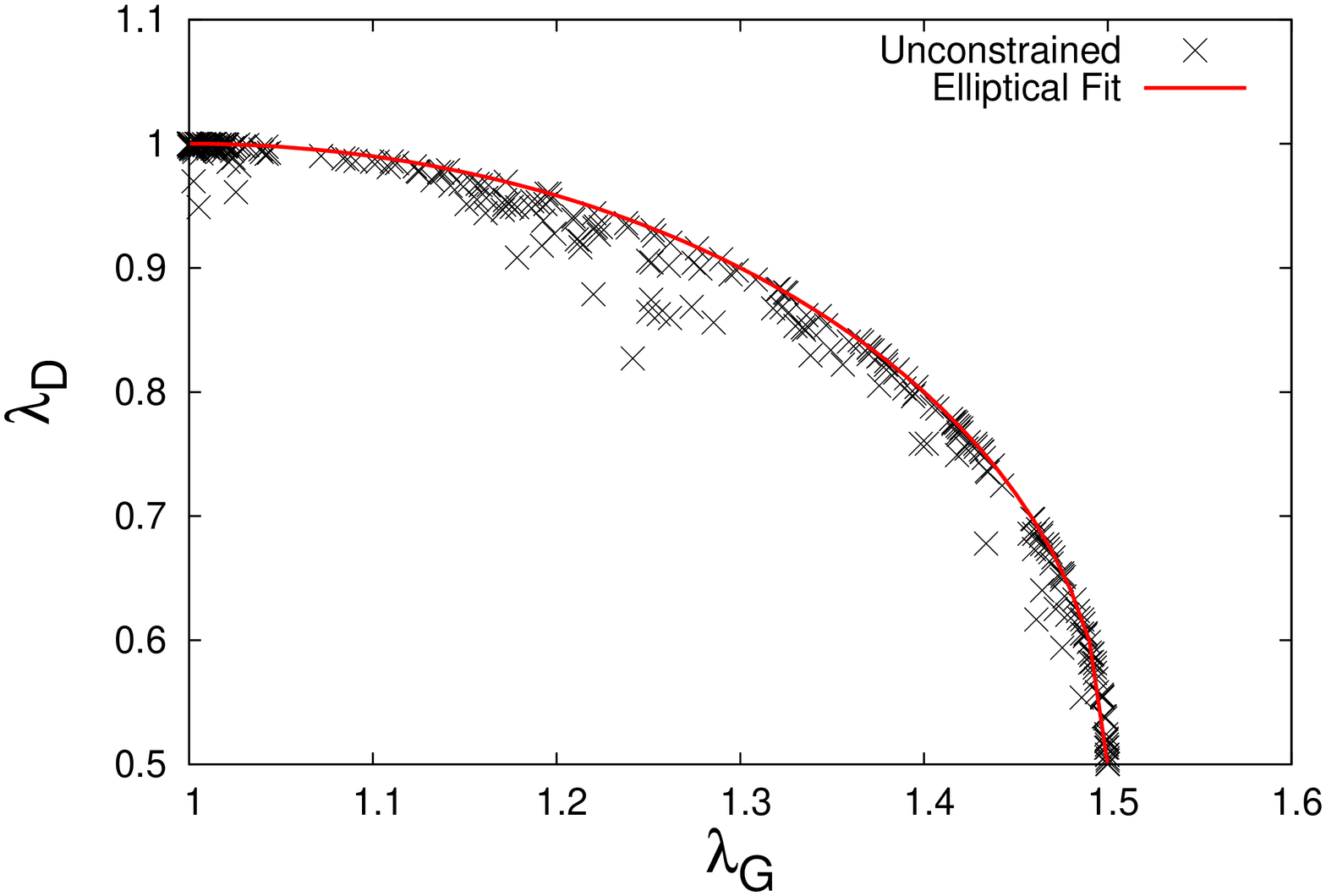}}\\
  \sidesubfloat[N=4]{\label{fig:N4}\includegraphics[scale=.32,angle=-90]{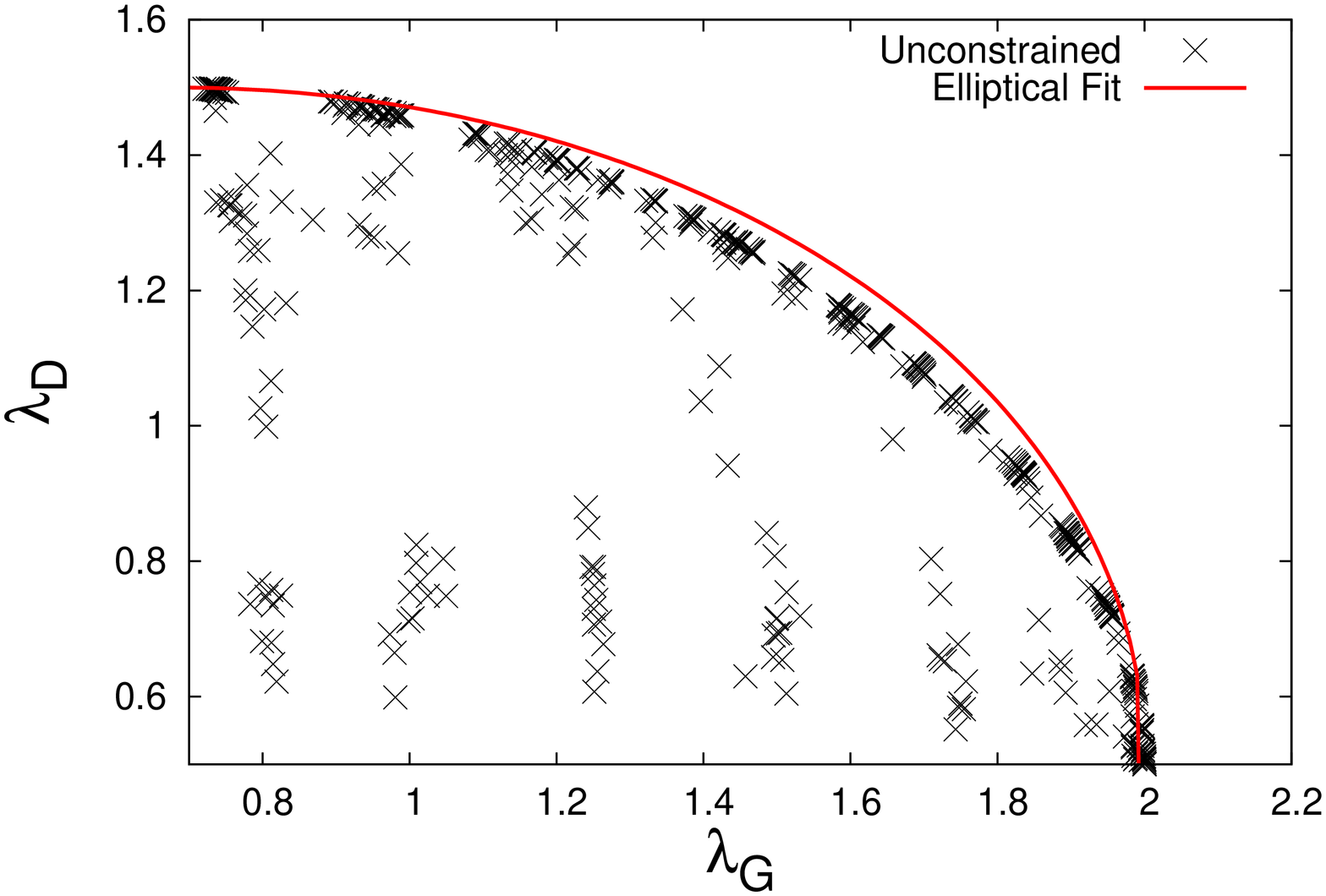}}\\
  \caption{Plots showing scans of $\lambda_D$  versus $\lambda_G$  are
	shown for unconstrained optimizations with a characteristic elliptical fit for the (a) $N=3$ and (b) $N=4$ cases.}\label{fig:Nplots}
\end{figure}

To visualize the entirety of $\lambda_D$ versus $\lambda_G$ space, we systematically varied the weight ($w_G$) and the specified eigenvalue of  $^2\tilde{G}$ ($\lambda_G^o$).  These visualizations for three and four electrons ($N=3$ and $N=4$) can be seen in Figs. \ref{fig:N3} and \ref{fig:N4}.
The $\lambda_D$ versus $\lambda_G$ space for each case was then fit with a characteristic ellipse that defines the maximum $\lambda_D$ for a given $\lambda_G$ and whose equation is given by
\begin{equation}
    \left(\frac{\lambda_D-\gamma_D^{\rm min}}{\gamma_D^{\rm max}-\gamma_D^{\rm min}} \right)^2+\left(\frac{\lambda_G
-\gamma_G^{\rm min}}{\gamma_G^{\rm max}-\gamma_G^{\rm min}} \right)^2=1
\label{eq:Ellipse}
\end{equation}
where the maximum eigenvalues of the $^2D$ and $^2\tilde{G}$ matrices are bounded  by $\lambda_D\in[\gamma_D^{\rm min},
\gamma_D^{\rm max}]$ and $\lambda_G\in[\gamma_G^{\rm min},\gamma_G^{\rm max}]$ respectively and where these maxima
($\gamma^{\rm max}$) and minima ($\gamma^{\rm min}$) were obtained from the previously-described scan over $\lambda _D$ versus $\lambda_G$ space.
\color{black}
Such elliptical nature of the boundary of the convex set of 2-RDMs when projected onto two dimensions has been
previously observed in the context of quantum phase transitions
[\onlinecite{schwerdtfeger_mazziotti_2009,gidofalvi_mazziotti_2006,zauner_2016}].
\color{black}

\color{black}

\color{black}

Fig. \ref{fig:N3} shows that for the case of three electrons in six orbitals ($N=3$, $r=6$),  the eigenvalues of the $^2D$ matrix lie in the range $\lambda_D\in[0.5,1]$, and the eigenvalues of the $^2\tilde{G}$ matrix lie in the range $\lambda_G\in[1,1.5]$.   As explained in the Theory section, strong fermion pair correlation is only seen when $\lambda_D$ exceeds one; hence, for the $N=3$ case, fermion pair condensation is not observed.  However, exciton condensation---seen when $\lambda_G$ exceeds one---can be obtained for this system.  In fact, maximum exciton condensation ($\lambda_G=\frac{3}{2}=1.5$) can be achieved.  Thus, while exciton and fermion pair condensation can not coexist for the $N=3$ case (as fermion pair condensation is not achievable), this small system can be used to further explore the properties of exciton condensation in future works.

\begin{figure}[tbh!]

\centering
\sidesubfloat[]{\label{fig:N4_optDG_test_b2}\includegraphics[width=6cm,angle=-90]{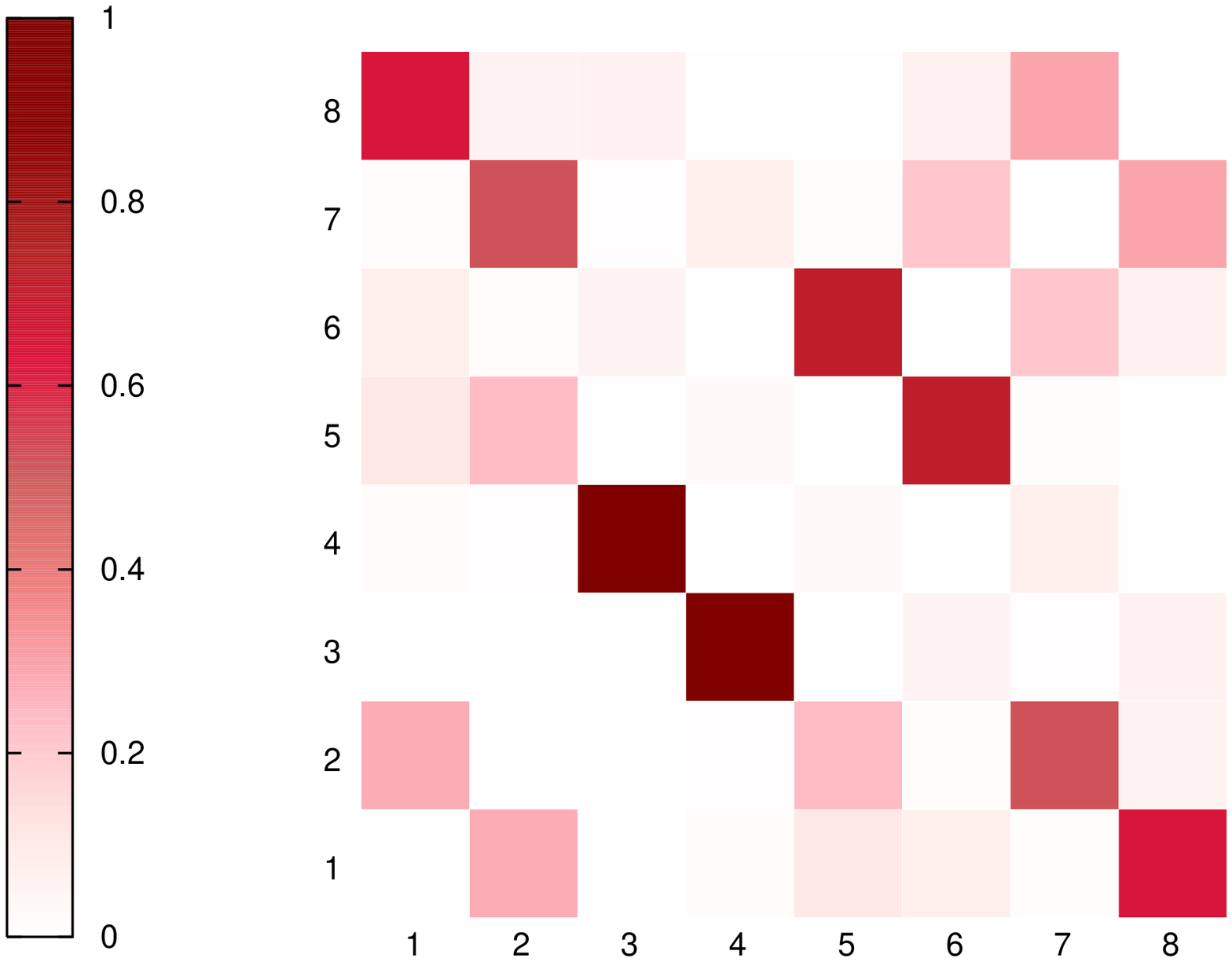}}\\
 \sidesubfloat[]{\label{fig:N4_optDG_test_b2G}\includegraphics[width=6cm,angle=-90]{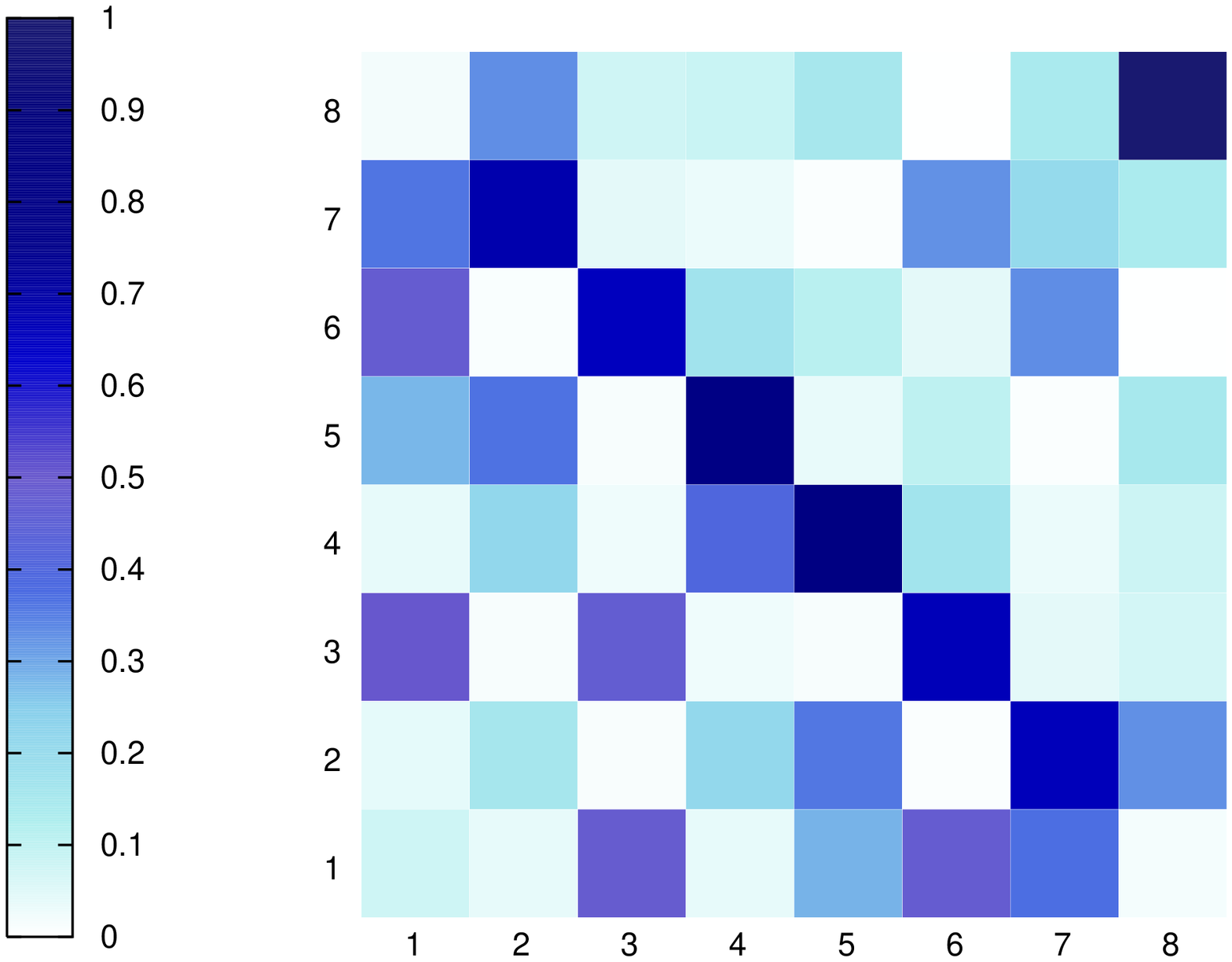}}
\caption{Visualizations of the (a) particle-particle pairs (red) and the (b) particle-hole pairs (blue) for optimizations of an $N=4$ calculation demonstrating both exciton and fermion pair condensation ($\lambda_G$=1.52, $\lambda_D=1.22$) are shown.  Note that the darker the shade of each color, the greater the extent of particle-particle/particle-hole pairing between orbitals. \color{black} The units of pair character are dimensionless.  \color{black}}
 \label{fig:N4comp}
\end{figure}

Fig. \ref{fig:N4} shows that for the case of four electrons in eight orbitals ($N=4$, $r=8$),  the eigenvalues of the $^2D$ matrix lie in the range $\lambda_D\in[0.5,1.5]$, and the eigenvalues of the $^2\tilde{G}$ matrix lie in the range $\lambda_G\in[0.7,2]$. Therefore, both excitonic and fermion pair condensation can be observed as both $\lambda_G$ and $\lambda_D$ exceed one for certain $N=4$ calculations.  Interestingly, there is indeed a region in $\lambda_D$ versus $\lambda_G$ space where both eigenvalues surpass one, demonstrating the simultaneous existence of exciton and fermion pair condensations.  However, as is apparent from the elliptical nature of the fit, there is a trade-off between the ability of the model system to exhibit high exciton character (a large $\lambda_G$) and high fermion pair character (a large $\lambda_D$).  This compromise between the two behaviors is shown schematically in Fig. \ref{fig:Tradeoff} and can be rationalized through analysis of the pairing behavior of the $r$ orbitals shown in Fig. \ref{fig:N4comp}.

Since excitons are particle-hole pairs and fermion pairs are particle-particle pairs, the existence of an excitonic relationship
between orbitals should preclude a fermionic relationship between the same orbitals and vice versa.
Figs. \ref{fig:N4_optDG_test_b2} and \ref{fig:N4_optDG_test_b2G} present the fermion-paired orbitals (red)
and the exciton-paired orbitals (blue)  respectively for an unconstrained $N=4$ calculation which demonstrated
simultaneous fermion pair and excitonic condensation ($\lambda_G$=1.52, $\lambda_D=1.22$).  Note that the
darker the shade of each color, the greater the extent of particle-particle (red) or particle-hole (blue) pairing
between the corresponding orbitals in the matrix plots.  As can be seen from Fig. \ref{fig:N4_optDG_test_b2}, the
orbital pairs with the largest particle-particle character are \{1,8\}, \{3,4\}, and \{5,6\}, and as can be seen from
Fig. \ref{fig:N4_optDG_test_b2G}, the orbital pairs with the largest particle-hole character are \{1,3\}, \{2,7\}, \{3,6\},
and \{4,5\}.  As expected, there is no overlap between strong fermion pair particle-particle orbital pairing and
strong exciton particle-hole orbital pairing.  These figures thereby confirm the apparent trade-off between exciton and
fermion pair character explicitly given by Eq. (\ref{eq:Ellipse}) and observed in Fig. \ref{fig:N4}.  Note that despite the trade-off,
neither the excitonic nor fermion pair character is trivial.  Both are delocalized across almost every pair of orbitals as can be seen
by the scarcity of white squares in Figs. \ref{fig:N4_optDG_test_b2} and \ref{fig:N4_optDG_test_b2G}.  As such, the coexistence
of fermion pair and excitonic character seems to be enmeshed in a significant manner.

\color{black}
\textbf{Large-$N$ Thermodynamic Limit---}To explore fermion-exciton condensation behavior in the thermodynamic limit ($N\rightarrow \infty$), we construct a general class of fermion-exciton-condensate wavefunctions $| \Psi_{FEC} \rangle$.  We first introduce model Hamiltonians known to exhibit fermion pair condensation and exciton condensation separately---the extreme Antisymmetrized Geminal Powers (AGP) Hamiltonian ($\hat{H}_{A}$) [\onlinecite{Richardson_1965,Coleman_1965,C1963}] and the Lipkin Hamiltonian ($\hat{H}_{L}$) [\onlinecite{Lipkin_model, texpansion, David1998, Stein_2000, David2004}], respectively.  As the extreme AGP model demonstrates maximal fermion condensation [\onlinecite{Coleman_1965}], there exists an eigenfunction of $\hat{H}_A$---$|\Psi_A\rangle$---whose largest eigenvalue of particle-particle RDM ($\lambda_D$) approaches the maximal limiting value
\begin{equation}
    Tr\left(\hat{d}\hat{d}^{\dagger}|\Psi_A\rangle\langle\Psi_A|\right)=\lambda_D^{(A)}= N
\label{eq:AGP}
\end{equation}
for systems of $2N$ or $2N+1$ fermions [\onlinecite{S1965,Coleman_1965}] where $\hat{d}$ is the operator of the eigenstate of ${}^{2}D$ corresponding to  $\lambda_D^{(A)}$.  Similarly, there exists an eigenfunction of the large-coupling $\hat{H}_L$---$|\Psi_L\rangle$---whose largest eigenvalue of the modified particle-hole RDM ($\lambda_G$) approaches the maximal limiting value
\begin{equation}
    Tr{\left ( \hat{g}\hat{g}^{\dagger}|\Psi_L\rangle\langle\Psi_L| \right )} =\lambda_G^{(L)}= \frac{N}{2}
\label{eq:Lip}
\end{equation}
for systems of $N$ fermions [\onlinecite{GR1969}] where $\hat{g}$ is the operator of the eigenstate of ${}^{2}\tilde{G}$ corresponding to $\lambda_G^{(L)}$.

Let the model wavefunction of the fermion-exciton condensate be given by the entanglement of the fermion-pair-condensate (AGP) and exciton-condensate (Lipkin) wavefunctions,
\begin{equation}
    |\Psi_{FEC}\rangle=\frac{1}{\sqrt{2 - |\Delta|}}\left(|\Psi_A\rangle - {\rm sgn}(\Delta) |\Psi_L\rangle \right),
\label{eq:FEC}
\end{equation}
in which $\Delta = 2\langle \Psi_A | \Psi_L \rangle$.  For this wavefunction, a lower bound on the largest eigenvalue of the particle-particle RDM is given by
\begin{align}
      \frac{1}{2} Tr\mathlarger{\mathlarger{\mathlarger{[}}}&\hat{d}\hat{d}^{\dagger}\mathlarger{\mathlarger{\mathlarger{(}}}|\Psi_A\rangle\langle\Psi_A|+|\Psi_L\rangle\langle\Psi_L| \nonumber\\
     & - 2~{\rm sgn}(\Delta)|\Psi_A\rangle\langle\Psi_L| \mathlarger{\mathlarger{\mathlarger{)]}}} \le \lambda_D^{(FEC)}.
\label{eq:eig_d}
\end{align}
From Eq. (\ref{eq:AGP}), the contribution to $\lambda_D^{(FEC)}$ from $|\Psi_A\rangle\langle\Psi_A|$ would be $\lambda_D^{(A)}$; additionally, as the extreme AGP model does not support exciton condensation, the contribution of $|\Psi_L\rangle\langle\Psi_L|$ to the eigenvalue must satisfy Pauli-like bounds of $0\le\gamma_D^{(L)}\le 1$ .  Limits on the cross terms can be obtained by representing the positive semidefinite matrix $\hat{d}\hat{d}^{\dagger}$ in the $\{|\Psi_A\rangle , |\Psi_L\rangle\}$ basis:
\begin{equation}
    \left( {\begin{array}{cc}
   \langle \Psi_A | \hat{d}\hat{d}^{\dagger} | \Psi_A\rangle & \langle \Psi_A | \hat{d}\hat{d}^{\dagger} | \Psi_L\rangle\\
   \langle \Psi_L | \hat{d}\hat{d}^{\dagger} | \Psi_A\rangle & \langle \Psi_L | \hat{d}\hat{d}^{\dagger} | \Psi_L\rangle \\
  \end{array} } \right)=
  \left( {\begin{array}{cc}
   \lambda_D^{(A)} & x\\
  x & \gamma_D^{(L)} \\
  \end{array} } \right)\succcurlyeq 0.
\label{eq:basis}
\end{equation}
As the determinant of this matrix must be greater than or equal to zero, the maximum contribution of the cross terms is
\begin{equation}
    \lambda_D^{(A)}\gamma_D^{(L)}-x^2\ge0\Rightarrow |x| \le \sqrt{\lambda_D^{(A)}\gamma_D^{(L)}}\le \sqrt{\lambda_D^{(A)}}.
\label{eq:cross}
\end{equation}
Inserting the lower-bound values into Eq. (\ref{eq:cross}) yields
\begin{equation}
    \lambda_D^{(FEC)}\ge \frac{1}{2}\lambda_D^{(A)}-\sqrt{\lambda_D^{(A)}}= \frac{N}{2}-\sqrt{N}
\label{eq:FEC_2}
\end{equation}
for 2$N$ or 2$N$+1 fermions.  Thus, as the number of fermions gets arbitrarily large ($N\rightarrow\infty$), $\lambda_D^{(FEC)}$ is simultaneously large.
Through an analogous derivation, it can be shown that a lower bound on the largest eigenvalue of the particle-hole RDM is given by
\begin{equation}
    \lambda_G^{(FEC)}\ge \frac{1}{2}\lambda_G^{(L)}-\sqrt{\lambda_G^{(L)}}= \frac{N}{4}-\sqrt{\frac{N}{2}}
\label{eq:FEC_3}
\end{equation}
for a system of $N$ fermions.  Thus, as the number of fermions gets arbitrarily large ($N\rightarrow\infty$), $\lambda_G^{(FEC)}$ is simultaneously large.  As an entanglement of the AGP and Lipkin wavefunctions demonstrates simultaneous large eigenvalues of ${}^{2}D$ and ${}^{2}\tilde{G}$, $|\Psi_{FEC}\rangle$ does indeed represent a fermion-exciton condensate in this large-$N$ thermodynamic limit.  \textcolor{black}{As superconducting systems exhibit fermion-pair condensation and as bilayer systems are known to exhibit exciton condensation, a possible experimental avenue for entangling the fermion-pair condensate ($|\Psi_A\rangle$) and exciton condensate ($|\Psi_L\rangle$) wavefunctions---and hence obtaining a fermion-exciton condensate---may be to construct a bilayer system composed of superconducting layers.}

\color{black}

\textit{Discussion and Conclusions:} In this study, we have theoretically observed the coexistence of both fermion pair and exciton condensation in a single quantum state: a fermion-exciton condensate. This concurrent character is not disparate; rather, the fermion pair and excitonic character are entwined in a highly non-trivial manner.  Still, there does appear to be an inherent trade-off between fermion pair and excitonic character following an elliptic relationship, which precludes the simultaneous presence of maximum fermion pair and maximum excitonic condensation.  However, as the number of electrons ($N$) is increased, the length of the major and minor axes of the ovular fit increase, causing the compromise between characters to become less and less stark.

\color{black}

We have also shown that a large class of fermion-exciton condensates can be constructed by entangling the wavefunctions of traditional fermion-pair condensates with exciton condensates.  Bounds on the large eigenvalues of their particle-particle and particle-hole RDMs establish the fermion-exciton condensates in the large-$N$ thermodynamic limit.  The entangled wavefunctions can be used for further theoretical and experimental exploration of properties.

\color{black}


A significant motivation for investigating fermion-exciton condensates is the possible
hybridization of the properties of both fermion pair condensates and exciton condensates.  A material that
combines the superconductive nature of fermion pair condensation [\onlinecite{BCS1957}, \onlinecite{Blatt_SC}]
with the dissipationless transport of energy of excitonic condensation [\onlinecite{KSE2004,TSH2004,Fil_Shevchenko_Rev}]
would have obvious applications in energy transport and electronics.  Now that the coexistence of fermion pair and excitonic
character in a fermion-exciton condensate has been computationally and theoretically established, further theoretical and
experimental studies\textcolor{black}{---possibly including studies on bilayer systems composed of superconductors---} are
needed.  There are certainly many open questions regarding the formation, the properties, the
applications, and the stability of fermion-exciton condensates that need to be explored in the following years.


\begin{acknowledgments}
\textit{Acknowledgments}: D.A.M.   gratefully   acknowledges   the   U.S.   National Science  Foundation  Grant No.  CHE-1565638
and  the  U.S.  Army Research Office (ARO) Grant No. W911NF-16-1-0152.
\end{acknowledgments}




\bibliography{references}
\end{document}